\documentclass[preprint,showpacs, preprintnumbers,amsmath,amssymb,nofootinbib]{revtex4}

\usepackage{graphicx}
\usepackage{dcolumn} 
\usepackage{bm}      

\newcommand{\bea}{\begin{eqnarray}}
\newcommand{\eea}{\end{eqnarray}}

\begin{document}
\title{Generalized Chaplygin gas model: constraints from Hubble parameter versus Redshift Data}         
\author{  Puxun Wu\;$^{a,b}$ and Hongwei Yu\;$^{a,}$\footnote{corresponding author}
}

\address
{   $ ^a$Department of Physics and Institute of  Physics,\\ Hunan
Normal University, Changsha, Hunan 410081, China
\\ $ ^{b}$School  of
Sciences and Institute of  Physics, Central South University of
Forestry and Technology, Changsha, Hunan 410004, China}


\begin{abstract}
We examine observational constraints on the generalized Chaplygin
gas (GCG) model for dark energy from the $9$ Hubble parameter data
points, the 115 SNLS Sne Ia data and the size of baryonic acoustic
oscillation peak at redshift, $z=0.35$. At a $95.4\%$ confidence
level, a combination of three data sets gives $0.67\leq A_s\leq
0.83$ and $-0.21\leq \alpha\leq 0.42$, which is within the allowed
parameters ranges of the GCG as a candidate of the unified dark
matter and dark energy. It is found that the standard Chaplygin gas
model ($\alpha=1$) is ruled out by these data at the $99.7\%$
confidence level.

\end{abstract}


 \maketitle

\section{Introduction}
Many astrophysical and cosmological observations, including  Type Ia
Supernovae (Sne Ia)\cite{Riess} and cosmic microwave background
radiation (CMBR)\cite{Balbi, WMAP}  etc, indicated that the universe
is undergoing an accelerating expansion. Many works have being done
in order to explain this discovery. Some people attribute the
observed acceleration to a possible breakdown of our understanding
of the laws of gravitation, thus they attempted to modify the
Friedmann equation \cite{Cardassian,DGP}. However, many more think
that the cosmic acceleration is driven by an exotic energy component
with the negative pressure in the universe, named dark energy, which
at late times dominates the total energy density of our universe and
accelerates the cosmic expansion. Up to now there are many
candidates of dark energy, such as the cosmological constant
$\Lambda$ \cite{Constant}, quintessence \cite{Quint}, phantom
\cite{Phantom} and quintom \cite{Quintom} etc.

Recently  an interesting model of dark energy,  named the
Chaplygin gas, was proposed  by Kamenshchik  et al
\cite{Kamenshchik}. This model is characterized  by an exotic
equation of state
 \bea \label{equationstate}
p_{ch}=-\frac{A}{\rho^\alpha_{ch}} \eea
 with a positive constant $A$
and $\alpha=1$. Progress has been made toward generalizing these
model parameters. In this regard, Bento et al. generalized
parameter $\alpha$ from $1$ to an arbitrary constant in Ref.
\cite{Bento2002}, and this generalized model was called the
generalized Chaplygin gas (GCG) model and can be obtained from a
generalized version of the Born-Infeld action. For $\alpha=0$ the
GCG model behaves like the scenario with cold dark matter plus a
cosmological constant.

Inserting the above equation of state of the GCG into the energy
conservation equation, it is easy to obtain
 \bea
 \rho_{ch}=\rho_{ch0}\bigg(A_s+\frac{1-A_s}{a^{3(1+\alpha)}}\bigg)^{\frac{1}{1+\alpha}}\;,
\eea
 where $\rho_{ch0}$ is the present energy density of the GCG and $A_s\equiv
 A/\rho^{1+\alpha}_{ch0}$.  It is worth noting that, when $0< A_s< 1$, the GCG model
smoothly interpolates between a non-relativistic matter phase
($\rho_{ch}\propto a^{-3}$) in the past and at late times a negative
pressure dark energy regime ($\rho_{ch}=-p_{ch}$). As a result of
this interesting feature,  the GCG model has been proposed as a
model of the unified dark matter and  dark energy (UDME). Meanwhile,
for $A_s=0$ the GCG behaves always like matter while for $A_s=1$ it
behaves always like a cosmological constant.

 The GCG model, thus,  has been the subject of great
interest and many authors have attempted to constrain this UDME
model by using various observational data, such as the Sne Ia
\cite{Fabris, Silva, Bean, Makler, Cunha, Zhu, Zhang}, the CMBR
\cite{Bean, Bento, CMBR}, the gamma-ray bursts \cite{GRB},  the
gravitational lensing \cite{Silva, Makler, lensing}, the X-ray gas
mass fraction of clusters \cite{Cunha, Makler, Zhu}, the large scale
structure \cite{Bean, large}, and the age of high-redshift objects
\cite{age}.

In this paper we shall consider the new observational constraints on
the parameter space of the GCG for a flat universe by using a
measurement of the Hubble parameter as a function of redshift
\cite{Jimenez}, the new 115 Sne Ia data released by the Supernova
Legacy Survey (SNLS) collaboration recently \cite{SNLS} and
 the baryonic acoustic oscillation (BAO) peak detected in the large-scale
correlation function of luminous red galaxies from Sloan Digital Sky
Survey (SDSS)\cite{SDSS}. We perform a combined analysis of three
databases and find that the degeneracy between $A_s$ and $\alpha$ is
broken. At a $95.4\%$ confidence level we obtain a strong constraint
on the GCG model parameters: $0.67\leq A_s\leq 0.83$ and $-0.21\leq
\alpha\leq 0.42$, a parameter range within which the GCG model could
be taken as a candidate of UDME and the pure Chaplygin gas model
could be ruled out.

\section{Constraint from the Hubble parameter as a function of redshift}
Last year, based on differential ages of passively evolving
galaxies determined from the Gemini Deep Deep Survey \cite{GDDS}
and archival data \cite{archival}, Simon et al. \cite{Simon} gave
an estimate for the Hubble parameter as a function of the redshift
$z$,
 \bea H(z)=-\frac{1}{1+z}\frac{dz}{dt}
 \eea
where $t$ is the time. They obtained $9$ data points of $H(z)$ at
redshift $z_i$ and used the estimated $H(z)$ to constrain the dark
energy potential. Later these $9$ data points were used to constrain
parameters of holographic dark energy model \cite{Yi}, parameters of
the $\Lambda CDM$, $XCDM$ and $\phi CDM$ models \cite{Samushia} and
the interacting dark energy models \cite{Wei}. Here we will use this
data to constrain the GCG model.

For a flat universe containing only the baryonic matter and the GCG,
the Friedmann equation can be expressed as
 \bea
 H^2(H_0, A_s, \alpha, z)=H_0^2 E^2(A_s, \alpha, z)\;,
 \eea
 where  \bea
E(A_s, \alpha,  z)=
[\Omega_b(1+z)^3+(1-\Omega_b)(A_s+(1-A_s)(1+z)^{3(1+\alpha)})^{\frac{1}{1+\alpha}}]^{1/2}\;,
 \eea
 $\Omega_b$ is the present dimensionless density parameter of baryonic  matter and
 $H_0=100hKms^{-1}Mpc^{-1}$ is present Hubble constant. The Hubble Space Telescope key projects give $h=0.72\pm
 0.08$ \cite{Freedman}  and the WMAP observations give
 $\Omega_bh^2=0.0233\pm0.0008$ \cite{WMAP}.  The best fit values for model parameters $
 A_s, \alpha$ and constant  $H_0$ can be determined by minimizing
 \bea
 \chi^2(H_0, A_s, \alpha)=\Sigma^9_{i=1}\frac{[H(H_0, A_s, \alpha,
 z_i)-H_{obs}(z_i)]^2}{\sigma^2(z_i)}\;.
 \eea
Since we are interested in the model parameters,  $H_0$ becomes a
nuisance parameter. We marginalize over $H_0$ to get the
probability distribution function of  $A_s$ and $\alpha$: $L(A_s,
\alpha)=\int dH_0P(H_0)e^{- \chi^2(H_0, A_s, \alpha)/2}$,  where
$P(H_0)$ is the prior distribution function for the present Hubble
constant. In this paper a Gaussian priors $H_0=72\pm 8 km S^{-1}
Mpc^{-1}$ is considered.

\begin{figure}[htbp]
\includegraphics[width=10cm]{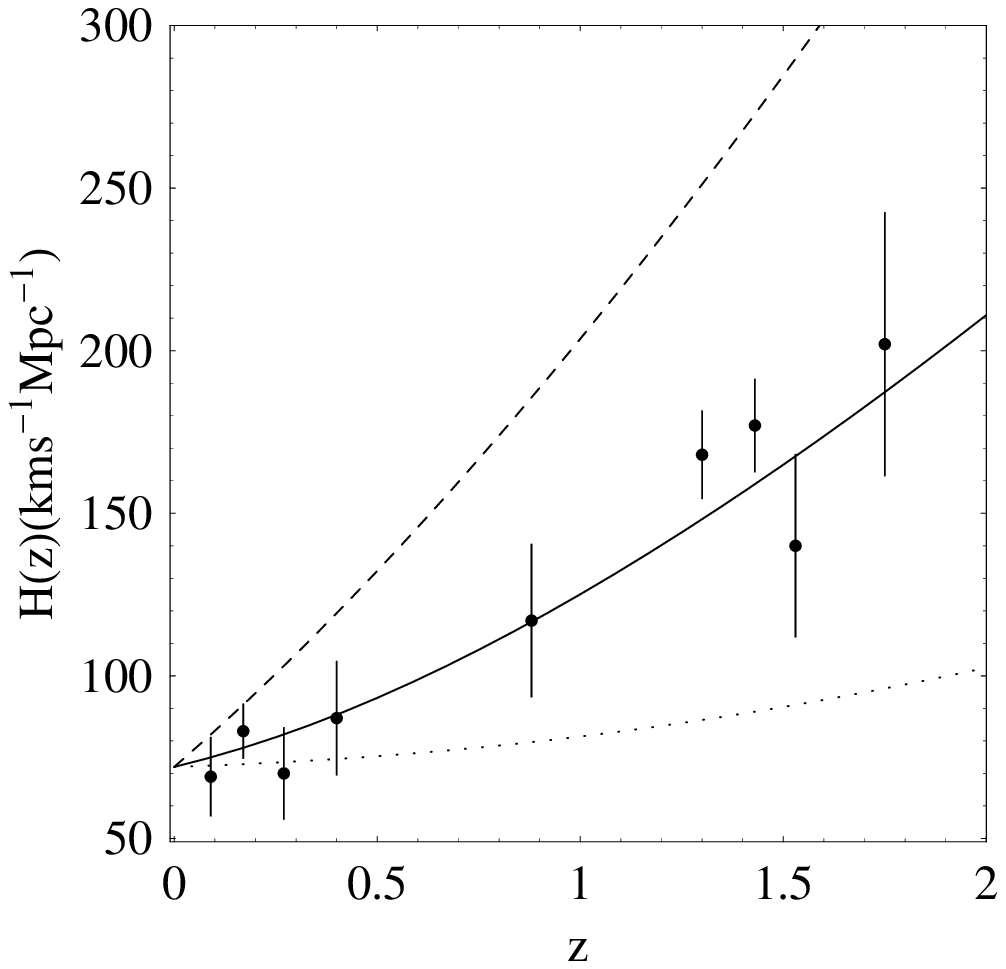}
\caption{\label{Fig1} The Hubble parameters $H(z)$  as a function of
$z$ for the case $H_0=72 kms^{-1}Mpc^{-1}$. The solid curve
corresponds to our best fit to 9 Hubble parameter data plus SNLS SNe
Ia data and SDSS baryonic acoustic oscillation peak with $A_s=0.75,
\alpha=0.05$. The dotted line and dashed  line correspond to
$A_s=1.0$ and $A_s=0.0$ respectively. }
\end{figure}

\begin{figure}[htbp]
\includegraphics[width=10cm]{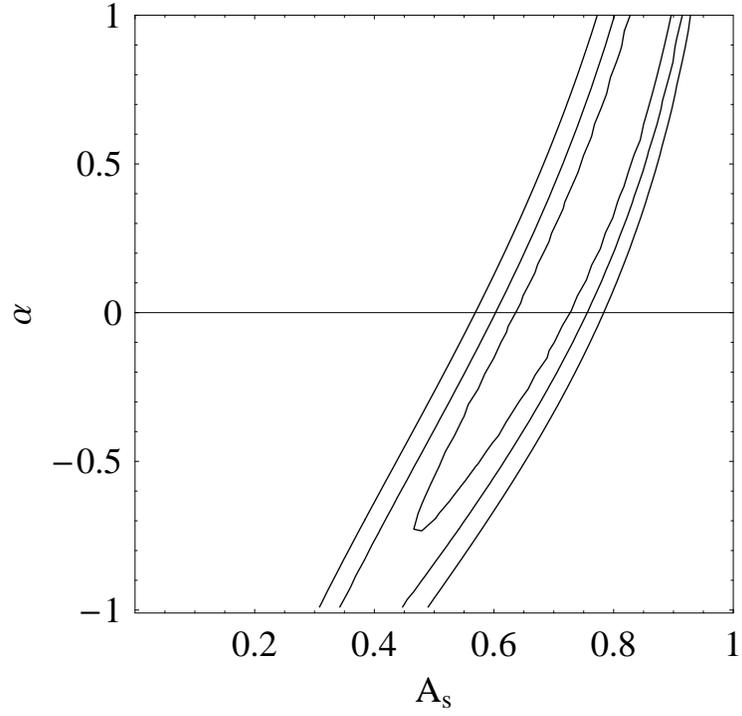}
\caption{\label{Fig2}  The $68.3\%$, $95.4\%$ and $99.7\%$
confidence level contours   for $A_s$ versus $\alpha$ from the
measurement of Hubble parameter with a  Gaussian priors $H_0=72\pm 8
km S^{-1} Mpc^{-1}$.  The best fit happens at $A_s=0.82$ and $
\alpha=0.71$. }
\end{figure}

In Fig. (\ref{Fig1}), we show the data of the Hubble parameter
plotted as a function of redshift for the case $H_0=72
kms^{-1}Mpc^{-1}$. Fig. (\ref{Fig2}) shows the results of our
statistical analysis for the Hubble parameter data. Confidence
contours (68.3\%, 95.4\% and 99.7\%) in the $A_s$-$\alpha$ plan
are displayed by considering the Hubble parameter measurements
discussed above.  The best fit happens at $A_s=0.82$ and
$\alpha=0.71$. It is very clear that  two model parameters, $A_s$
and $\alpha$, are degenerate.

\section{Joint statistics with SDSS BAO and SNLS Sne Ia}
\begin{figure}[htbp]
\includegraphics[width=10cm]{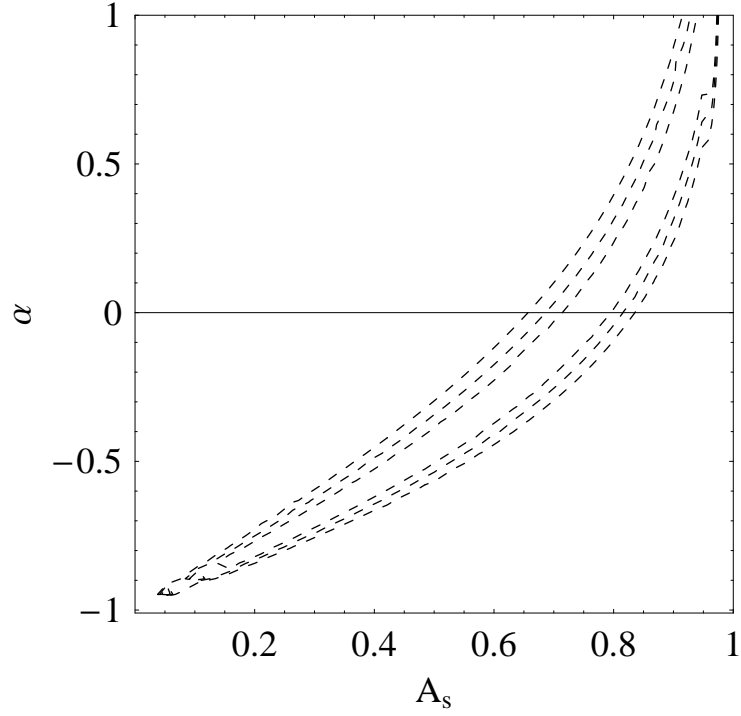} 
\caption{\label{Fig3} The $68.3\%$, $95.4\%$ and $99.7\%$ confidence
level contours  for $A_s$ versus $\alpha$  from the SDSS baryonic
acoustic oscillations.  The best fit happens at $A_s=0.76$ and $
\alpha=0.01$.}
\end{figure}

Using a large spectroscopic sample of 46,748 luminous red galaxy
from the SDSS, last year Eisenstein et al \cite{SDSS} successfully
found the size of baryonic acoustic oscillation (BAO) peak and
obtained a parameter $\mathcal{A}$, which is independent of
cosmological models and for a flat universe can be expressed as
\begin{eqnarray}
\mathcal{A}=\frac{\sqrt{\Omega_m}}{E(z_1)^{1/3}}\bigg[\frac{1}{z_1}
  \int_0^{z_1}\frac{dz}{E(z)}\bigg]^{2/3}\;,
\end{eqnarray}
 where $z_1 = 0.35$,  $\mathcal{A}$ is measured to be $\mathcal{A} = 0.469\pm 0.017$ and  $\Omega_m$ is the effective
matter density parameter given by
$\Omega_m=\Omega_b+(1-\Omega_b)(1-A_s)^{1/(1+\alpha)}$
 \cite{Makler, Zhu, Lima}.
Using  parameter $\mathcal{A}$ we can obtain the constraint on dark
energy models from the BAO. In Fig. (\ref{Fig3}) we show the
constraints from this measurement on the parameter space
$A_s-\alpha$. The best fit happens at $A_s=0.76$ and $\alpha=0.01$.
Although the BAO data constrains efficiently the parameter plane
into a narrow strip, parameters $A_s$ and $\alpha$ are also
degenerate.

\begin{figure}[htbp]
\includegraphics[width=10cm]{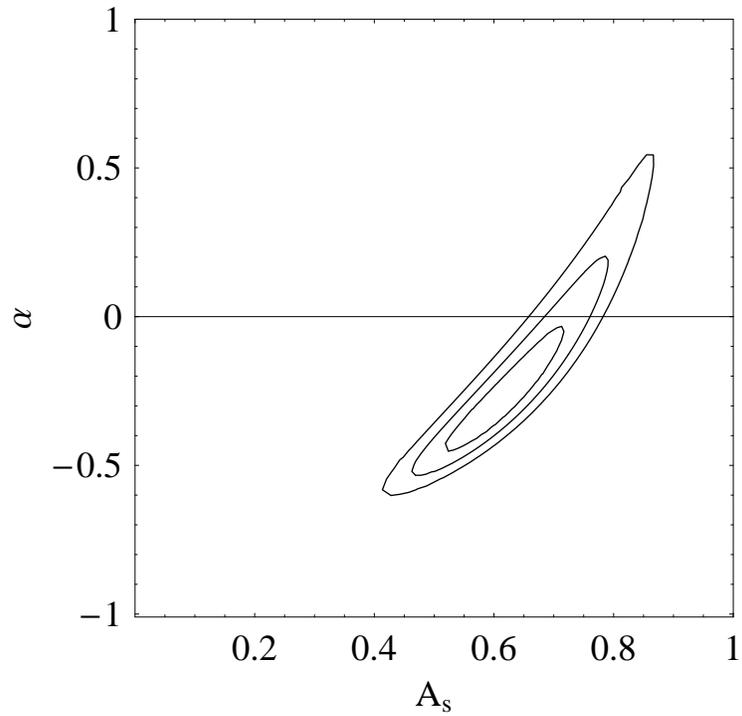} 
\caption{\label{Fig4} The $68.3\%$, $95.4\%$ and $99.7\%$ confidence
level contours for $A_s$ versus $\alpha$  from the Hubble parameter
data plus  the SDSS baryonic acoustic oscillations peak. The best
fit happens at $A_s=0.61$ and $ \alpha=-0.28$.}
\end{figure}

However, from Fig. (\ref{Fig2}, \ref{Fig3}) it is interesting to see
that possible degeneracies between these parameters may be broken by
combining these two kinds of observational data. In Fig.
(\ref{Fig4}) we show the results of such an analysis. The best fit
happens at $A_s = 0.61$ and $\alpha = -0.28$. At the $95.4\%$
confidence level we obtain $0.46\leq A_s \leq 0.79$ and $-0.53 \leq
\alpha \leq 0.2$, a stringent constraint on the GCG. Apparently at
the $68\%$ confidence level the scenario of  standard dark energy
plus dark matter scenario (i.e. the case of $\alpha=0$) is excluded.

\begin{figure}[htbp]
\includegraphics[width=10cm]{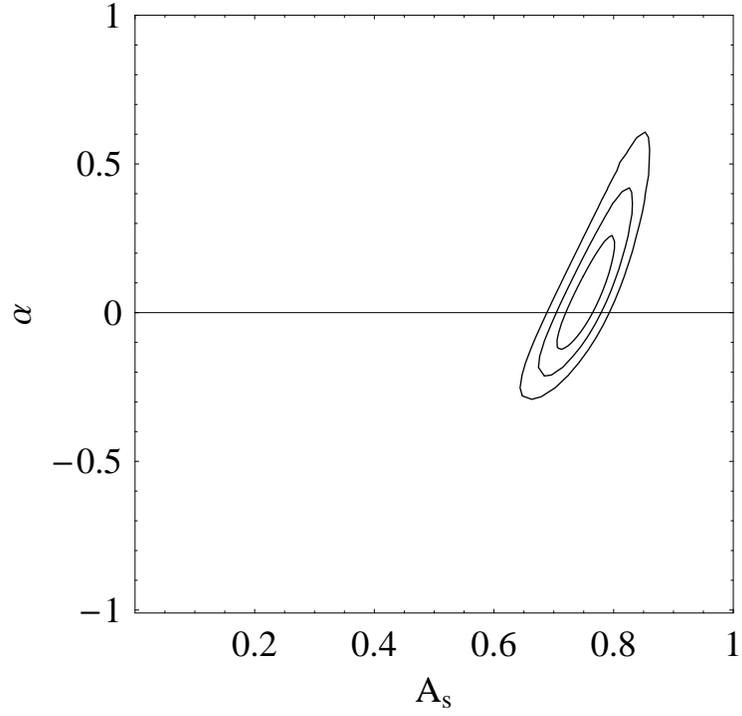} 
\caption{\label{Fig5} The $68.3\%$, $95.4\%$ and $99.7\%$ confidence
level contours for $A_s$ versus $\alpha$  from the Hubble parameter
data plus  the SDSS baryonic acoustic oscillations peak and the SNLS
Sne Ia data. The best fit happens at $A_s=0.75$ and $ \alpha=0.05$.}
\end{figure}

If further adding the new 115 SNLS Sne Ia data~\cite{SNLS}, which
contains  44 previously published nearby Sne Ia ($0.015<z<0.125$)
plus 71 distant Sne Ia ($0.15<z<1$) discovered by SNLS and gives the
best fit values, $A_s=0.78$ and $\alpha=0.16$, for the GCG model, we
find that a more stringent constraint is obtained, namely, at the
$95.4\%$ confidence level a combination of three databases gives
 $0.67\leq A_s\leq 0.83$ and $-0.21\leq \alpha\leq
0.42$ with the best fits $A_s=0.75$ and $\alpha=0.05$. In
Fig.(\ref{Fig5}) we show the $68.3\%$, $95.4\%$ and $99.7\%$
confidence level contours from these three data sets. It is easy
to see that our results are consistent with the standard dark
energy plus dark matter scenario at a $68\%$ confidence level.

\section{conclusion and discussion}
The constraints on the generalized Chaplygin gas (GCG) model,
proposed as a candidate of the unified dark matter-dark energy
scenario (UDME), has been studied in this paper. The Hubble
parameter as a function of redshift has been used to constrain the
parameter space of the GCG model. We find, although the Hubble
parameter gives a degeneracy between model parameters $A_s$ and
$\alpha$, the complementary and interesting constraints on the
parameters of the model could be obtained. Combining the new SNLS
Sne Ia data and the recent measurements of the baryon acoustic
oscillations found in the SDSS, we obtained a very stringent
constraint on model parameters of GCG. At the $95.4\%$ confidence
level, we found  $0.67\leq A_s\leq 0.83$ and $-0.21\leq \alpha\leq
0.42$. At addition we find at a $68\%$ confidence level the
combination of these three databases allows the scenario of standard
dark energy plus dark matter, although the Hubble parameter plus the
SDSS BAO exclude it.

Using the X-ray gas mass fractions of galaxy clusters and the
dimensionless coordinate distance of Sne Ia and FRIIb radio
galaxies, Zhu \cite{Zhu} obtained, at a $95.4\%$ confidence level,
$A_s=0.70_{-0.17}^{+0.17}$ and $\alpha=-0.09_{-0.33}^{+0.54}$. Using
the CMBR power spectrum measurements from BOOMERANG and Archeops,
together with the Sne Ia constraints, Bento et al. \cite{Bento}
found that $0.74<A_s<0.85$, and $\alpha<0.6$. Apparently these
results are comparable with our results in this paper,  which are
within the allowed parameters ranges of the GCG as a candidate of
UDME. However the standard Chaplygin gas model ($\alpha=1$) is ruled
out by these data at the $99.7\%$ confidence level. Meanwhile it is
easy to see that at a $68.3\%$ confidence level our result is
consistent with the standard dark energy plus dark matter scenario
(i.e. the case of $\alpha=0$), which is also in agreement with what
obtained in Ref.\cite{Zhu, Bento}.

\begin{acknowledgments}
 We would like to thank Z. Zhu for his valuable discussions and help.  This work was supported in part by the National
Natural Science Foundation of China  under Grants No. 10375023 and
No. 10575035, the Program for NCET under Grant No. 04-0784 and the
Key Project of Chinese Ministry of Education (No. 205110).
\end{acknowledgments}

\end{document}